\documentstyle[twoside,fleqn,npb,epsfig]{article}
\newcommand{\sla}{\kern -5.4pt /}
\newcommand{\slalarge}{\kern -10 pt /}
\newcommand{\Dir}{\kern -6.4pt\Big{/}}
\newcommand{\Dirin}{\kern -10.4pt\Big{/}\kern 4.4pt}
\newcommand{\DDir}{\kern -7.6pt\Big{/}}
\newcommand{\DGir}{\kern -6.0pt\Big{/}}

\newcommand{\be}{\begin{equation}}
\newcommand{\ee}{\end{equation}}
\newcommand{\bea}{\begin{eqnarray}}
\newcommand{\eea}{\end{eqnarray}}
\newcommand{\beanon}{\begin{eqnarray*}}
\newcommand{\eeanon}{\end{eqnarray*}}
\newcommand{\ba}{\begin{array}}
\newcommand{\ea}{\end{array}}
\newcommand{\bd}{\begin{description}}
\newcommand{\ed}{\end{description}}
\newcommand{\bt}{\begin{tabular}}
\newcommand{\et}{\end{tabular}}
\newcommand{\bi}{\begin{itemize}}
\newcommand{\ei}{\end{itemize}}
\newcommand{\ben}{\begin{enumerate}}
\newcommand{\een}{\end{enumerate}}
\newcommand{\bc}{\begin{center}}
\newcommand{\ec}{\end{center}}
\newcommand{\bflr}{\begin{flushright}}
\newcommand{\eflr}{\end{flushright}}

\newcommand{\bfll}{\begin{flushleft}}
\newcommand{\efll}{\end{flushleft}}

\newcommand{\OO}{{\it O}}

\newcommand{\ar}{\rightarrow}

\newcommand{\wph}{{\tt WPHACT}}

\newcommand{\wto}{{\tt WTO}}

\newcommand{\wb}{W}
\newcommand{\GeV}{\;\mathrm{GeV}}

\newcommand{\proccctwenty}{$e^-e^+\to e^-\bar{\nu}_e u\bar{d}$}
\newcommand{\proccceighteen}{$e^-e^+\to e^-\bar{\nu}_e \mu^+  \nu_{\mu}$}
\newcommand{\procmixfiftysix}{$e^-e^+\to e^-e^+ \nu_e\bar{\nu}_e$}
\newcommand{\procccten}{$e^-e^+\to \mu^-\bar{\nu}_\mu u\bar{d}$}

\newcommand{\gbs}{g^2}
\newcommand{\ssW}{{\scriptscriptstyle{\wb}}}
\newcommand{\sW}{p_{_W}}
\newcommand{\Ptg}{\Pi_{_{3Q}}}

\def\epem{\ifmmode{e^+ e^-} \else{$e^+ e^-$} \fi}

\def\pl #1 #2 #3 {{ Phys.~Lett.} {\bf#1} (#2) #3}
\def\np #1 #2 #3 {{ Nucl.~Phys.} {\bf#1} (#2) #3}
\def\zp #1 #2 #3 {{ Z.~Phys.} {\bf#1} (#2) #3}
\def\pr #1 #2 #3 {{ Phys.~Rev.} {\bf#1} (#2) #3}
\def\prep #1 #2 #3 {{ Phys.~Rep.} {\bf#1} (#2) #3}
\def\prl #1 #2 #3 {{ Phys.~Rev.~Lett.} {\bf#1} (#2) #3}
\def\intj #1 #2 #3 {{ Int. J. Mod. Phys.} {\bf#1} (#2) #3}
\def\mpl #1 #2 #3 {{ Mod.~Phys.~Lett.} {\bf#1} (#2) #3}
\def\rmp #1 #2 #3 {{ Rev. Mod. Phys.} {\bf#1} (#2) #3}
\def\cpc #1 #2 #3 {{ Comp. Phys. Commun.} {\bf#1} (#2) #3}
\def\xx #1 #2 #3 {{\bf#1}, (#2) #3}

\title{Fermion Loops, conserved currents and single-W}

\author{ Alessandro Ballestrero \address {I.N.F.N. and Dipartimento di Fisica 
Teorica, Torino}
}

\begin {document}
\begin {abstract}
The relevance of fermion loop corrections to four fermion processes 
at \epem colliders is reviewed with regard to the recent extension to
the case of  massive external particles  and its application
to single-W processes.
\end{abstract}

\maketitle

\section { Introduction} 

The problem of preserving gauge invariance in the  calculations involving 
unstable particles is well known since several years 
\cite{bhf1,baurzepp,lopez,bhf2} . It is connected with
the fact that the use of a width (e.g. $i\Gamma M$) in the denominator of an
$s$-channel unstable boson propagator is necessary to prevent divergences at 
$p^2=M^2$, but  it is in fact an effective way
of including only a part of  higher order corrections, and it therefore
violates gauge invariance.
  
There are various examples in which this violation becomes  numerically 
relevant.  For instance, this happens at tree level with four fermion final 
states for WW production at high energies, where gauge cancellations among 
the three double resonant (CC03) diagrams become relevant, and for
contributions at low $e^-$ angle. 
This last case is relevant for final states like \proccctwenty , 
\proccceighteen\ or \procmixfiftysix\
with $e^-$ undetected, which are commonly referred to as  single-W processes.
For them,  gauge invariance determines the 
 behaviour of the amplitude as a function of $t$. 

In the above mentioned  examples even small violations of  gauge invariance
may easily give results which are completely unreliable, and differ from
the correct ones not by a few percent but by some large factor.

Various gauge restoring methods have been described in the literature and have 
been used to avoid such inconsistencies. Some of the most used are : 

\begin{description}

\item[{\em Fixed width(FW):}]  In all massive-boson propagators one performs 
everywhere the substitution
$ M^2\; \ar\;  M^2-i\Gamma M $.
   This gives an unphysical width in $t$-channel, but retains  U(1) gauge 
   invariance.
\item[{\em Complex Mass(CM):}]
The substitution
$M^2\; \ar\;  M^2-i\Gamma M $ is applied  not only in propagators but also in 
relations involving couplings. This gives  unphysical complex couplings in 
addition to the width in $t$-channel.
It has however the advantage of   preserving both U(1) and SU(2) Ward identities.
\item[{\em Overall scheme(OA):}]
    When resonant propagators are present, all diagrams  (not only 
     resonant ones) are multiplied by 
    $(q^2-M^2)/(q^2-M^2+iM\Gamma)$. This method retains  gauge invariance but  
mistreats  non resonant terms.

\item[{\em Fermion Loop(FL):}] 
        At least the imaginary part of  all (propagators and vertices) 
   fermion loop corrections is used together with resummed boson propagators.
  The real part of FL corrections can as well be considered. It is not
  necessary for gauge restoration, but it 
     constitutes an important  gauge  invariant  subset of the radiative 
   corrections, which automatically determines the correct evolution of the 
    coupling constants.

\end{description}

   It has to be noticed that the first schemes  are somehow
   ``ad hoc'' prescriptions  to restore gauge invariance  
     introducing some non correct feature, whose consequences are expected
  to be of  little numerical impact but this should in principle be verified 
  case  by case. 
     FL  on the contrary is the only   fully consistent and 
    justified scheme in field theory. 
 
In the following we will review the FL scheme and its application to single-W  
processes, with particular regard to the
  latest results which  take into 
account current non conservation (massive external fermions).

\section { Fermion Loop  } 

The application of FL corrections  to 4f processes has been fully described
in the papers of ref. \cite{bhf1,bhf2}  

In the first paper only the Imaginary part of FL corrections (IFL) have been 
considered and as a case study the process  
 \proccctwenty\ has been used.
The approximation of considering massless
external fermions has been taken, which implies that all external
current are conserved (CC).

In this approximation U(1) Ward Identities are satisfied by adding a 
fixed width $i\Gamma M$ to all denominators of W propagators 
(also in $t$-channel), even if there is no 
 physical justification for this procedure. 

IFL corresponds to  resumming the  imaginary  part of fermionic loop 
corrections in propagators and adding  the imaginary part of vertex loop 
corrections. This has to be considered as the 
minimum set of corrections that has to be added in order to restore 
gauge invariance. 

With massless  fermions in the internal loops and in the  decay width, 
$i\frac{\Gamma p^2}{M}$ in propagators satisfies gauge invariance.

Some  numerical applications for \proccctwenty\ have been considered in 
\cite{bhf1}.

The complete FL corrections have instead been computed in ref \cite{bhf2}. 
Again in massless external fermion approximation, but with
massive  fermions in loops.
The renormalization of gauge boson masses at their (gauge invariant) 
complex poles 
\cite{Veltman} has been used.
It turns out that in this scheme 
all corrections can be reabsorbed in running couplings and renormalized 
propagator functions and triple vertices, so that an effective Born 
prescription can be used in 
which only tree level type diagrams  appear.

Numerical applications to \proccctwenty\ at low $e^-$ angle 
and \procccten\ at high energies show differences from other schemes.
At high energies differences also with IFL are found.

\section { Single W  } 

Processes with at least one electron and one electron-neutrino like 
 \proccctwenty\,  $e^-e^+\to e^-\bar{\nu}_e \mu^+  \nu_\mu$,
and   $e^-e^+\to e^- e^+ \bar{\nu}_e \nu_e$,
besides being relevant to WW or ZZ physics, are particularly interesting
in the kinematical configuration in which
the electron is  lost in the pipe. In such a configuration (single-W), 
they become important  as a background to searches and
for anomalous coupling studies.
 The cross sections are significant because of $t$-channel 
contributions, and they are directly measured at LEP2.

Single-W processes are  divergent in massless external fermion 
approximation. Therefore not only external electron masses have to be exactly
accounted for, but also those of the other  fermions ($u$, $d$,
$\mu$,..). 

Several fully massive 4fermion MonteCarlo's  are now available:
{\tt COMPHEP\cite{com}, GRC4F\cite{grc}, WPHACT\cite{wph}, KORALW\cite{kor}, 
NEXTCALIBUR\cite{nextc}, SWAP\cite{swap} }and   {\tt WTO}\cite{wto} 
which accounts for masses where they become important. 

\begin{figure}[htb]
  \begin{center}
  \unitlength 1cm
  \begin{picture}(4.,5.)
  \put(-2.,-2.5){\includegraphics{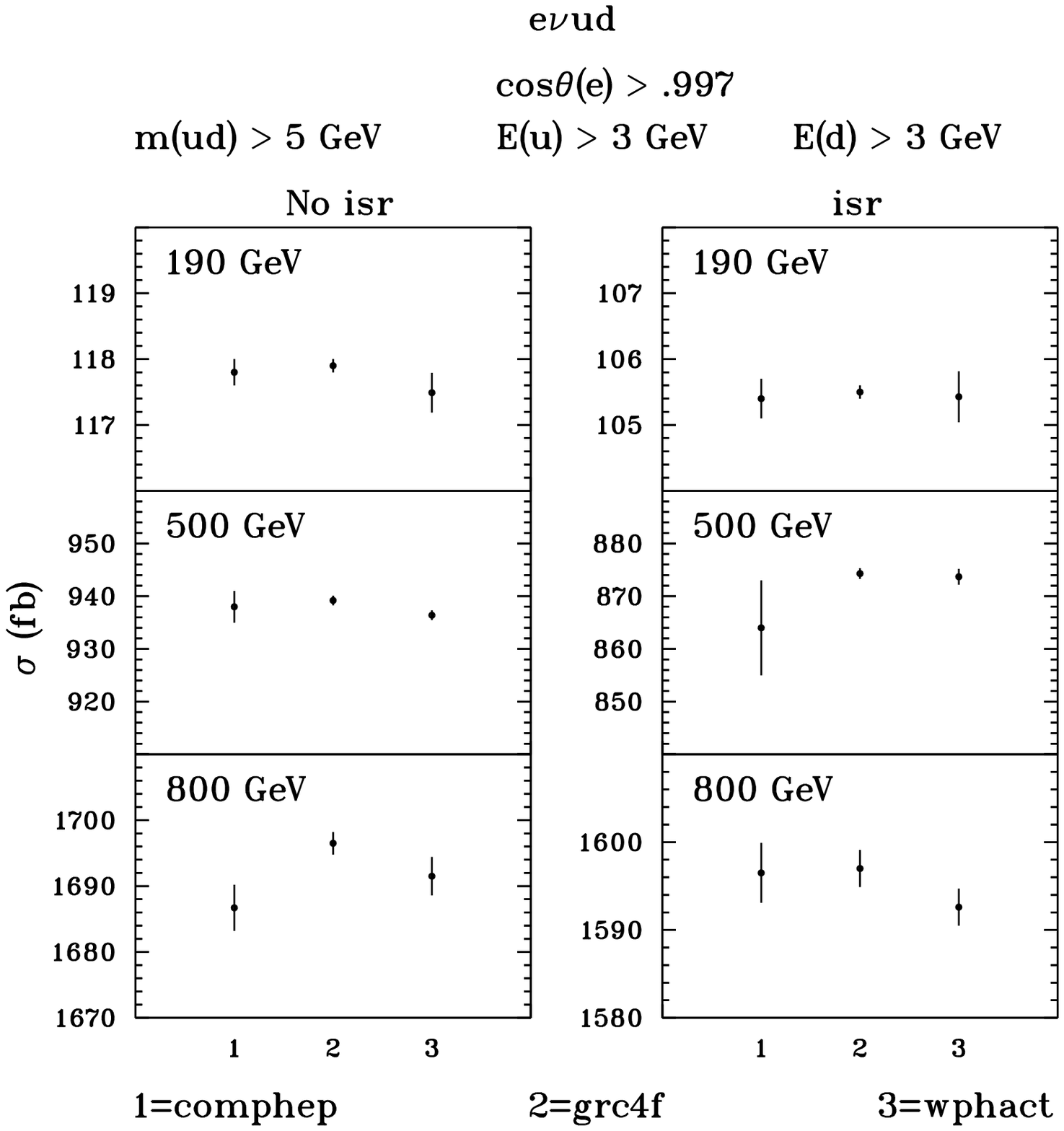}}
  \end{picture}
  \end{center}
  \caption {Comparison of the results from three  codes for the 
single-W process \proccctwenty .}
  \label{swcomparison}
\end{figure}

Good technical agreement among all these codes has been achieved for
single-W processes in tuned comparisons. In fig.~\ref{swcomparison} one
can find the results from the first three codes up to Linear Collider
energies, while we refer to Ref.~\cite{YR4f} for further comparisons with the
others at LEP2 energies.
It has to be remarked that in fig.~\ref{swcomparison} the three codes
 used  respectively 
overall,   $L_{\mu\nu}$ transform method \cite{lmunu} and fixed width as gauge
restoring schemes.    

\section { IFL and Non Conserved Currents  } 

FL calculations of ref.~\cite{bhf1,bhf2}  are not appropriate for single-W
processes because of the assumption of conserved currents 
(massless external fermions), which is in conflict with the necessity
to account for external fermion masses.

Nevertheless, numerical studies have  been performed using these results 
together with fully massive matrix elements in \cite{Hoogland&vanoldenborgh},
where it was noticed that the corresponding U(1) gauge  violation
is  proportional to $m_e$. It can however be enhanced
by large factors at high energy, as it will be shown in table~\ref{tabschemes} .

The Imaginary Fermion Loop scheme  with fully massive ME and exact non 
conserved current contributions has been studied recently~\cite{ifl}
and implemented in \wph.

The unitary gauge has been used.
While  for massless internal fermions the Ward Identities (WI) are satisfied 
by the fixed width propagator 

\[  {-i \over p^2- M^2 +i\Gamma M_W}
        \left\{
         g^{\mu\nu} 
         - {{p^\mu p^\nu} \over {M^2-i\Gamma M_W}}
               \right\}, 
\]

the correct resummed propagator is instead

\[  {-i \over p^2- M^2 +i\Pi}
        \left\{
         g^{\mu\nu} 
         - {p^\mu p^\nu \over M^2}
               \left(1 + {i\Pi \over p^2}\right)\right\} 
\]

with ``running width'' $\Pi=p^2 \frac{\Gamma_W}{M_W}$

With it, WI are  properly satisfied  only if exact IFL triple vertex 
corrections are computed.

As it can be deduced from tables \ref{tabschemes} ,  \ref{tabfwifl},  
the use of IFL does not give  significant numerical 
differences with ``ad hoc'' schemes  for cross sections with typical 
single-W cuts. The last line of  table~\ref{tabschemes} shows instead that the  
 approximation of  conserved currents together with massive ME 
lead to inconsistent numerical results at high energies.

\begin{table}[htb]
\setlength{\tabcolsep}{.4pc}
\caption{ Cross sections for  $e^+ e^- \to e^- \bar \nu_e u\bar d$ 
and various gauge restoring schemes, all implemented in \wph .
No ISR.
 $M(u \bar d) > 5 \GeV\; , E_u > 3 \GeV, E_{\bar d} > 3 \GeV,
\cos (\theta_e ) > .997$ }
\label{tabschemes}
\vskip .1cm
\begin{tabular}{ c c c c } 
\hline
            & 190 GeV         & 800 GeV         & 1500 GeV           \\
\hline
IFL       &   .11815 (13)   &   1.6978 (15)   &    3.0414 (35)     \\
FW        &   .11798 (11)   &   1.6948 (12)   &    3.0453 (41)     \\
CM        &   .11791 (12)  &  1.6953 (16)  &  3.0529  (60)   \\
OA        &   .11760 (10)   &   1.6953 (13)   &    3.0401 (23)     \\   
\hline
IFLCC     &   .11813 (12)   &   1.7987 (16)   &    5.0706 (44)     \\
\hline
\end{tabular}
\end{table}

\begin{table}[htb]
\setlength{\tabcolsep}{.4pc}
\caption{Comparison of FW and IFL schemes for different single-W
cross sections at $\sqrt{s} = 200$ GeV. No ISR.
 $|\cos\theta_e| > 0.997$, $E_{\mu} > 15\,$GeV, 
and $|\cos\theta_{\mu}| < 0.95$.
First line $M_{u\bar{d}}> 45$ GeV, second line $M_{u\bar{d}}< 45$ GeV }
\label{tabfwifl}
\vskip .1cm
\begin{tabular}{ c c c } 
\hline
Final state         & IFL        & FW          \\
\hline
$e^-\bar{\nu}_e u\bar{d}$   
      &   0.12043 (10)   &   0.12041 (11)       \\
$e^-\bar{\nu}_e u\bar{d}$ 
      &   0.028585 (14)   &   0.028564 (14)       \\
$e^-\bar{\nu}_e \mu^+ \nu_\mu $
      &   0.035926 (34)   &   0.035886 (32)   \\
$e^-\bar{\nu}_e e^+ \nu_e $
      &   0.050209 (38)  &  0.050145 (32)   \\
\hline
\end{tabular}
\end{table}

\begin{figure}[htb]
  \begin{center}
  \unitlength 1cm
  \begin{picture}(4.,6.)
  \put(-1.5,-1.8){\includegraphics{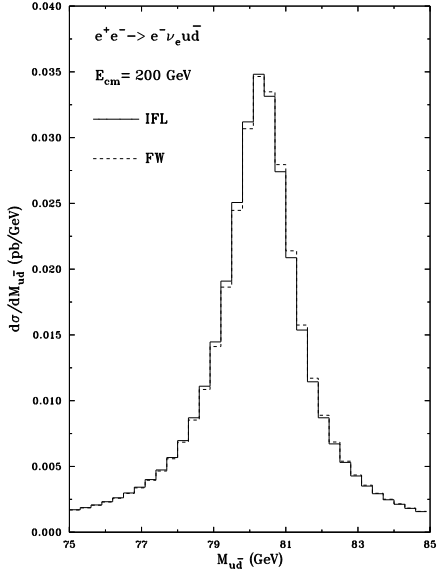}}
  \end{picture}
  \end{center}
  \caption {Mass distribution of the $u\bar d$ pair computed by \wph\  for
fixed with and imaginary fermion loop schemes.
 No ISR. $|\cos \theta_e^-| > 0.997$} 
  \label{massdistr}
\end{figure}

Some differences between IFL and other schemes is evident 
(fig~\ref{massdistr}) when
one considers mass distributions. This is connected to the fact that
with IFL one properly makes use of the running width.

\section { FL and Non Conserved Currents  } 

Big theoretical uncertainties for single-W processes, in absence of complete 
\OO ($\alpha$) corrections, are connected to the scales of the couplings  and 
to the scale for ISR with dominating $t$-channel contributions.

Complete FL calculations with massive external fermions have been recently
performed\cite{fl}.
These are necessary to solve  the first of the uncertainties just mentioned. 

 Complex mass renormalization has been used as in \cite{bhf2} and it leads 
again to  effective Born calculations. 
Besides running couplings and renormalized propagators 
now also running boson  masses are needed. They are  defined by
\[
\frac{1}{M^2(p^2)} = \frac{1}{M^2}\,{{p^2-S^0_{\ssW}+\frac{M^2}{p^2}\,S_{\phi}}
\over {p^2-S_{\phi}}}
\]

which for a massless internal world reduces to
\[
M^2(p^2) = \frac{g^2(p^2)}{g^2(\sW)}\,\sW, \qquad M^2(\sW) = \sW.
\]

\[
\frac{1}{\gbs(s)} = \frac{1}{\gbs} - \frac{1}{16\,\pi^2}\,\Ptg(s)
\]

It has to be remarked that, using  the Feynman gauge, complete one loop 
resummation gives a W-propagator which is  equivalent  to 
some effective  ``unitary gauge'' form:

\[
\Delta^{\mu\nu}_{\rm eff} = \frac{1}{p^2-M^2+S^0_{\ssW}}\,\Big[
\delta^{\mu\nu} + \frac{p^{\mu}p^{\nu}}{M^2(p^2)}\Big].
\]

The complete FL scheme for non conserved currents has been implemented in 
\wto\ for \proccctwenty\ 
and  $e^-e^+\to e^-\bar{\nu}_e \mu^+ \bar \nu_\mu$, where the fermion masses 
are  accounted for ``when needed'':  in $t$-channel 
$\gamma$ exchange 
diagrams for $log(m^2/s)$ and constant contributions.

\begin{figure}[htb]
  \begin{center}
  \unitlength 1cm
  \begin{picture}(4.,6.5)
  \put(-2.,-2.){\includegraphics{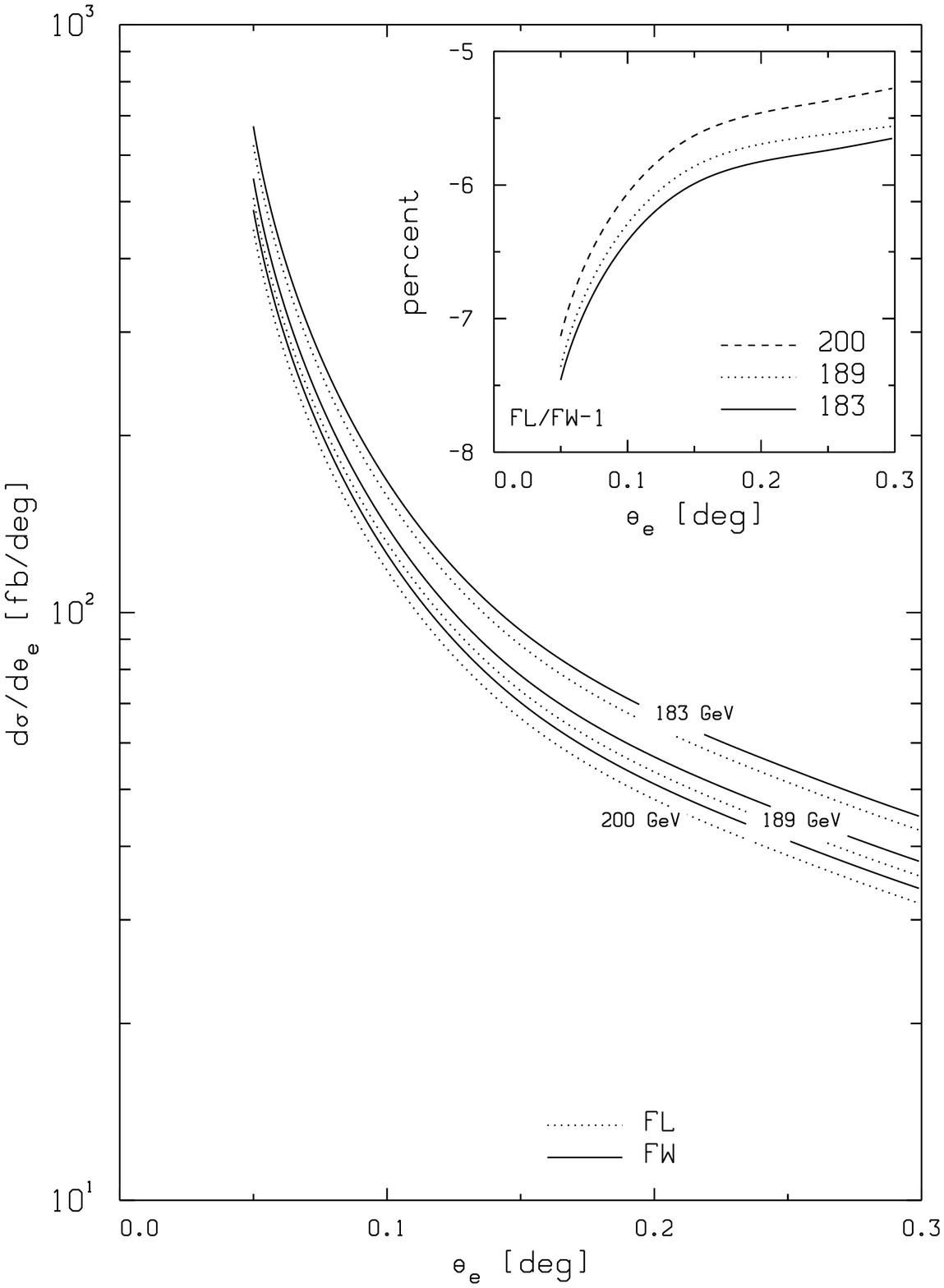}}
  \end{picture}
  \ec
  \caption{$d\sigma /d \theta_e$ distribution computed by \wto\ for 
\proccctwenty\ and $M(u\bar d) > 45$ GeV}
  \label{gpfig}
\end{figure}

The numerical results~\cite{fl} show  differences with tree 
level fixed width $G_f$  scheme up to $\sim 7\%$. 
This result is reported in fig~\ref{gpfig}.


One may expect that the bulk of such a  difference comes from $\alpha$ running 
effects, as it is rather obvious that the value of $\alpha_{em}$ in $G_f$ 
renormalization scheme is not the correct one to describe $t$-channel $\gamma$
propagator 
couplings. For such a reason,  $\alpha (t)$ with IFL (IFL$\alpha$) 
has been implemented in \wph\  for  $t$-channel contribution only.
 
For cuts used at LEP2 energies for single-W, this seems to describe with very 
good approximation  \proccctwenty , as can be deduced by a comparison with
FL both for cross sections (table~\ref{sechad})
and for angular distributions (table~\ref{anghad}).

\begin{table}[htb]
 \caption{Total single-$W$ cross-section in fb for  $e^+e^- \to e^- \bar
  \nu_e u \bar d$, for $M(u\bar d) > 45\,$GeV and $|\cos\theta_e| > 0.997$.
  No ISR.
   FL is computed by \wto, IFL and IFL$\alpha$ by \wph. }
 \label{sechad}
\vskip .1cm
 \begin{tabular}{c c c c}
 \hline
 $\sqrt{s}$  & IFL &IFL$\alpha$ & FL  \\
 \hline
 $183\,\GeV$  & 88.50(4)   & 83.26(5) & 83.28(6)    \\
 $189\,\GeV$  & 99.26(4)   & 93.60(9) & 93.79(7)    \\
 $200\,\GeV$  & 120.43(10) & 113.24(8)& 113.67(8)   \\
 \hline
 \end{tabular}
\end{table}

\begin{table}[htb]
 \caption{$d\sigma/d\theta_e$  [pb/degrees] for  $e^+e^- \to e^- \bar\nu_e 
u \bar d$. $M(u\bar d) > 45\,$GeV, $\sqrt s$=200 GeV. No ISR.
FL is computed by \wto, IFL and IFL$\alpha$ by \wph.
}
 \label{anghad}
\vskip .1cm
 \begin{tabular}{c c c c }
 \hline
 $\theta_e\,$[Deg]  & IFL &IFL$\alpha$ & FL   \\
 \hline
$0.0^\circ \div 0.1^\circ$ 
    & 0.67077    & 0.62404   & 0.62357   \\
$0.1^\circ \div 0.2^\circ$ 
    & 0.09321     & 0.08753   & 0.08798     \\
$0.2^\circ \div 0.3^\circ$ 
    & 0.05455    & 0.05141   & 0.05141     \\
$0.3^\circ \div 0.4^\circ$ 
    & 0.03867    & 0.03624   & 0.03646     \\
 \hline
 \end{tabular}
\end{table}

\begin{figure}[htb]
  \begin{center}
  \unitlength 1cm
  \begin{picture}(4.,6.)
  \put(-2.2,-1.5){\includegraphics{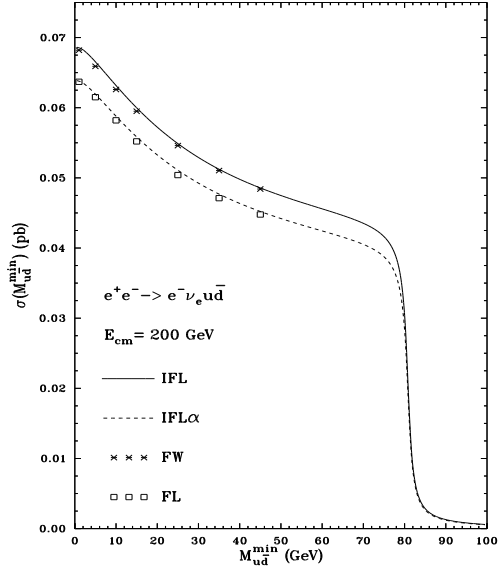}}
  \end{picture}
  \ec
  \caption{Total cross section for $\theta_e\,<\,0.1^\circ$ as a function of the lower 
cut on M$_{u\bar d}$ in IFL and IFL$\alpha$ 
schemes.  Curves  computed with \wph. Markers give the results of 
FW and FL by \wto . }
  \label{fliflmas}	
\end{figure}

Moreover, it has been checked in fig.~\ref{fliflmas} that 
the agreement does not depend on the $M(u\bar d)$ invariant mass cut
and it remains below 1\% 
down to the very low cuts.

For the process $e^-e^+\to e^-\bar{\nu}_e \mu^+  \nu_\mu$ 
instead, the  difference between
FL and  IFL$\alpha$ turns out to be  of the order of 2-3\% both in cross 
sections (table~\ref{seclep}) and in angular distributions 
(table~\ref{anglep}). The reason for such a different behaviour between the
two processes is probably due to the cuts and the 
relative importance of multiperipheral contributions in them.  
The results of tables~\ref{seclep},\ref{anglep} in any case indicate that
the running of $\alpha$ is not in general sufficient for a very  accurate 
description of the effects accounted for by complete FL calculations

\begin{table}[htb]
 \caption{Total single-$W$ cross-section in fb for $e^+e^- \to e^- \bar 
\nu_e \mu^+ \nu_{\mu}$, for $|\cos\theta_e| > 0.997$, $E_{\mu} > 15\,$GeV, 
and $|\cos\theta_{\mu}| < 0.95$. No ISR. 
 FL is computed by \wto, IFL and IFL$\alpha$ by \wph. }
 \label{seclep}
\vskip .1cm
 \begin{tabular}{c c c c}
 \hline
 $\sqrt{s}$  & IFL &IFL$\alpha$ & FL  \\
 \hline
 $183\,\GeV$ &26.45(1)  & 24.90(1)  & 25.53(4)     \\
 $189\,\GeV$ &29.70(2)  & 27.98(2)  & 28.78(4)       \\
 $200\,\GeV$ &35.93(4)  & 33.85(4)  & 34.97(6)      \\
 \hline
 \end{tabular}
\end{table}

\begin{table}[htb]
 \caption{$d\sigma/d\theta_e$ in [pb/degrees] for  $e^+e^- \to e^- \bar\nu_e 
\nu_{\mu} \mu^+$, for $|\cos\theta_e| > 0.997$, 
$E_{\mu} > 15\,$GeV, and $|\cos\theta_{\mu}| < 0.95$. 
$\sqrt{s} = 183\,$GeV. No ISR.
 FL is computed by \wto, IFL and IFL$\alpha$ by \wph.
}
 \label{anglep}
\vskip .1cm
 \begin{tabular}{ c c c c}
 \hline
 $\theta_e\,$[Deg]  & IFL &IFL$\alpha$ & FL \\
 \hline
$0.0^\circ \div 0.1^\circ$ 
            &0.14170  &0.1319 & 0.13448                     \\
$0.1^\circ \div 0.2^\circ$ 
            &0.02117  &0.01987 & 0.02031               \\
$0.2^\circ \div 0.3^\circ$ 
            &0.01240  &0.01166 & 0.01194                     \\
$0.3^\circ \div 0.4^\circ$ 
            &  0.00879  & 0.00830  & 0.00851           \\
 \hline
 \end{tabular}
\end{table}

\begin{figure}[htb]
  \begin{center}
  \unitlength 1cm
  \begin{picture}(4.,6.5)
  \put(-2.2,-1.5){\includegraphics{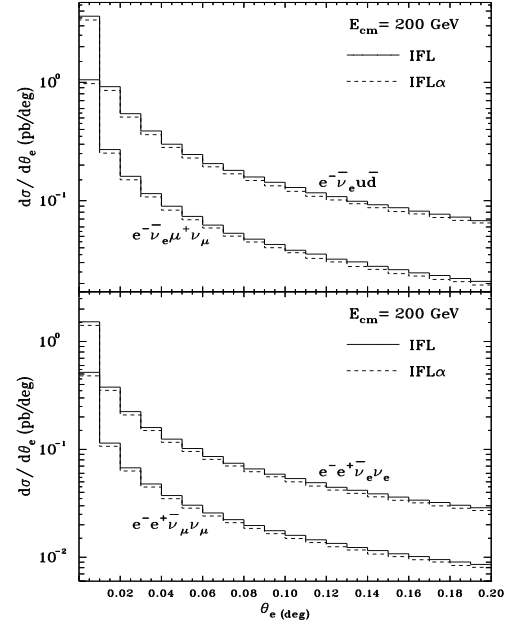}}
  \put(12.5,21.5){\wph}
  \end{picture}
  \ec
  \caption{Angular distributions for different single-W processes at
$\sqrt{s}$ = 200 GeV in the IFL and IFL$\alpha$ scheme.  }
  \label{swall}
\end{figure}

In fig.~\ref{swall} one can see the differences of IFL and IFL$\alpha$ 
angular distributions for various processes, including 
$e^+e^- \ar e^+e^-\nu \bar \nu$ where FL corrections are not available yet.

For the latter, the estimate by \wph\ of the theoretical uncertainty of 
IFL$\alpha$ 
calculations  is  of the order of about
3\% \cite{YR4f}. This refers only to the uncertainty connected to the
absence of complete FL calculations and not to the one due to 
the treatment of ISR/FSR in presence of dominant $t$-channel 
contributions~\cite{YR4f}.

\section { Conclusions} 

FL corrections have been extended to massive external fermions  in 4f physics.
With them calculations can be safely performed down to $\theta_e =0$.

Single-W processes represent one of the most important applications of such
extension.

IFL corrections show that ``ad hoc'' gauge restoring schemes are reliable for 
total cross sections but may produce some differences in distributions.

Complete FL are a gauge invariant subset of radiative corrections, essential
for single-W processes.

Their results differ from $G_f$ scheme by several percent.

IFL + $\alpha$ running for $t$-channel reproduces FL results at less than 
1\% level
for \proccctwenty\ and at 2-3\% for 
$e^-e^+\to e^-\bar{\nu}_e \mu^+  \nu_{\mu}$.

Further theoretical uncertainties for single-W, connected with ISR in $t$-channel
dominated processes have not been considered in this short account. 

Further analyses and improvements are probably still needed for single-W
processes, expecially at Linear Collider.

\section*{Acknowledgments} I wish to thank  Johannes Bl\"umlein and 
Tord Riemann for the pleasant  and constructive atmosphere 
at the 5th Zeuthen Workshop ``Loops and Legs in Quantum Field Theory''.

I am particularly grateful to Elena Accomando and Ezio Maina for their 
collaboration.

\end{document}